\documentclass[draftcls, 12pt,onecolumn,oneside,compress]{IEEEtran}
\usepackage{times,amsmath,color,amssymb,graphicx,epsfig,cite,psfrag,subfigure,algorithm,balance}
\usepackage{amsfonts,pifont,enumerate,cases}
\usepackage{mathrsfs}
\usepackage[table]{xcolor}
\usepackage{verbatim}
\usepackage{geometry}
\usepackage{bm}
\newtheorem{property}{Property}

\newtheorem{remark}{\underline{Remark}}

\newtheorem{proof}{Proof}

\newcommand{\tu}{\textup}

\begin{document}
\title{Time Varying Channel Tracking with Spatial and~Temporal~BEM~for~Massive~MIMO~Systems}
\author{Jianwei~Zhao, Hongxiang Xie,
        Feifei~Gao,
        ~Weimin~Jia, Shi Jin, and Hai Lin 
\thanks{This work was presented in part at the 2017 IEEE ICC \cite{ICC}.}
\thanks{J. Zhao, H. Xie and F. Gao are with  Tsinghua National Laboratory for Information Science and Technology (TNList) Beijing 100084, P. R. China, (e-mail:zhaojw15@mails.tsinghua.edu.cn, xiehx14@mails.tsinghua.edu.cn, feifeigao@ieee.org). J. Zhao is also with
High-Tech Institute of Xi’an, Xi’an, Shaanxi 710025, China.}
 \thanks{ W.~Jia is with High-Tech Institute of Xi’an, Xi’an, Shaanxi 710025, China (e-mail: jwm602@163.com).}
 \thanks{ S.~Jin is with the National Communications Research Laboratory, Southeast University, Nanjing 210096, P. R. China (email: jinshi@seu.edu.cn). }
 \thanks{H. Lin is with the Department of Electrical and Information Systems,
Graduate School of Engineering, Osaka Prefecture University, Sakai,
Osaka, Japan (e-mail: hai.lin@ieee.org).}
\vspace{-10mm}}

\maketitle
\begin{abstract}
In this paper, we propose a channel tracking method for massive MIMO systems under both time-varying and spatial-varying circumstance. Exploiting the characteristics of massive antenna array, a spatial-temporal basis expansion model (ST-BEM) is designed to reduce the effective dimensions of uplink/downlink channel, which decomposes channel state information (CSI) into the time-varying spatial information and gain information. We firstly model the users' movements as a one-order unknown Markov process, which is blindly learned by the expectation and maximization (EM) approach. Then, the uplink time-varying spatial information can be blindly tracked by unscented Kalman filter~(UKF) and  Taylor series expansion of the steering vector, while the rest uplink channel gain information can be trained by only a few pilot symbols. Due to angle reciprocity (spatial reciprocity), the spatial information of the downlink channel can be immediately obtained from the uplink counterpart, which greatly reduces the complexity of downlink channel tracking. Various numerical results are provided to demonstrate the effectiveness of the proposed method.
\end{abstract}

\begin{IEEEkeywords}
\vspace{-4mm}
Massive MIMO, channel tracking,  spatial and temporal BEM (ST-BEM), DOA, angle reciprocity, unscented Kalman Filter (UKF).
\end{IEEEkeywords}
\IEEEpeerreviewmaketitle
\section{Introduction}
 Massive multiple-input multiple-output (MIMO), as one of the most important techniques in 5G communications, has attracted enormous attention from both academy and industry \cite{marzetta2010noncooperative,chen1,chen2}. It applies hundreds or even thousands of antennas at the base station (BS) to simultaneously serve tens of terminals in the same time-frequency resource, making the communications much more efficient, secure, and robust\cite{five, han(9),scalingupMIMO}.

The potential gains of massive MIMO depend on the channel state information (CSI), and numerous works have been proposed to solve channel estimation issues \cite{xie,xieCR,overview,Caire,yin,KLT,CCS}. In most cases, the massive array antennas are adopted at base station (BS), which makes the elements of channel possess high correlation in the spatial domain, and the corresponding channel covariance appears to be low rank. Based on this fact,  \cite{Caire} proposed to use dominant eigenvectors to span the channel vector for the time-invariant circumstance and reduce the effective channel parameters. The compressive sensing (CS) technique was proposed to reduce the training as well as the feedback overhead in \cite{KLT,CCS}, while a novel pilot decontamination approach was presented in \cite{yin} that offers a powerful way of discriminating across interfering users with even strongly correlated pilot sequences. Recently, the angle domain massive MIMO channel estimation scheme were proposed in \cite{xie,xieCR,overview,zhao,linhai,fandian}, where the angle information is exploited to separate users, and the array signal processing method is utilized to simplify the subsequent design, also named as angle division multiple access~(ADMA). The works in \cite{xie,xieCR,overview,zhao,linhai,fandian} all utilized the fact that the angle information could be distinguished precisely for massive MIMO systems with large number of antennas, by contrast, which would be not valid for the traditional MIMO systems, such as the smart antenna. Therefore, massive MIMO is an effective juncture of the wireless communications and the array signal processing.

Nevertheless, a practical propagation of wireless signals would face the time-varying environment. The first work considering  channel aging effect on massive MIMO was \cite{channelaging1}, where the channel variation was characterized as a function of different system parameters, based on which a channel prediction method  was designed. In \cite{channelaging2}, the authors investigated the impact of the general channel aging conditions on the downlink performance of massive MIMO systems, while reference \cite{channelaging3} discussed the effect of channel aging on the sum rate of uplink massive MIMO systems. However, most of these works   focus on the performance analysis rather than providing a concrete approach  for  dynamic channel tracking.

Conventionally, three approaches can be used for channel aging problem: (1) the Gauss-Markov model \cite{gao_relay14}, which captures the channel temporal variation through the symbol-by-symbol updating; (2) the basis expansion model (BEM) \cite{GiannakisBEM}, which decomposes the channels into the superposition of the time-varying basis functions weighted by the time-invariant coefficients; (3) the temporal channel covariance matrix method, where the dominant eigenvectors of the temporal covariance matrix serve as the basis vectors to span the time-varying channels. However, there are no related channel tracking methods for large-scale massive MIMO systems to the best of our knowledge.

In this paper, we propose a channel tracking method for massive MIMO systems under both the time-varying and the spatial-varying circumstance. A spatial-temporal basis expansion model (ST-BEM) is designed to reduce the effective dimensions of uplink/downlink channels, which decomposes channel state information (CSI) into the time-varying spatial information and gain information. Besides, the spatial information can be further determined by the central direction of arrival (DOA) and angular spread (AS) of the incoming signal. We firstly model the users' movements as a one-order Markov process whose unknown parameters are estimated by expectation and maximization (EM) learning. Then, the central DOAs can be blindly tracked by unscented Kalman Filter (UKF), while the AS can also be blindly obtained through the Taylor series expansion of the steering
vector. With the tracked spatial information, the uplink channel gain can be obtained by only a few pilot symbols. Besides, due to angle reciprocity, the downlink channel spatial information can be immediately obtained from uplink counterpart, which greatly reduces the complexity of downlink channel tracking. Finally, various numerical results are provided to demonstrate the effectiveness
of the proposed method. Compared to the conference version \cite{ICC}, this journal version has more contributions as:
\begin{itemize}
  \item The users' movements are modeled as a one-order unknown Markov process which is a more practical case of mobile users.
  \item The DOAs of the incident signals are assumed as time varying and are tracked by EM learning based UKF.
  \item The ADMA scheme is utilized to decrease the user interference and improve the spectrum efficiency.
  \item Various numerical results are provided to verify the effectiveness of the proposed method.
\end{itemize}

The rest of the paper is organized as follows. In section \ref{sec:2}, the system model and the channel model are given.
The problem formulation of channel tracking is described in section \ref{sec:3}. Spatial signature tracking is discussed in section \ref{sec:4}, while channel gain estimation is presented in section~\ref{sec:5}. Numerical simulations and results are displayed in section \ref{sec:6}. Finally, conclusions are drawn in section \ref{sec:7}.

\textbf{Notations:} Vector is denoted by boldface small letter, while matrix is expressed by
capital letter;  ${\mathbf A}^T$, ${\mathbf A}^*$, ${\mathbf A}^H$, ${\mathbf
A}^{-1}$ and ${\mathbf A}^\dag$ represents the transpose, complex conjugate, hermitian,
inverse, and pseudo-inverse of the matrix ${\mathbf A}$ respectively; the trace of ${\mathbf A}$ is ${\rm tr}({\mathbf
A})$; the $(i,j)$th entry of ${\mathbf A}$ is $[{\mathbf A}]_{ij}$; a diagonal matrix
with the diagonal elements constructed from ${\mathbf a}$ is denoted by diag$\{{\mathbf a}\}$, while diag$\{\mathbf A\}$
is a vector with elements extracted from the diagonal components of matrices $\mathbf A$;
$\mathbb{E}\{\cdot\}$ is the statistical
expectation and $\left\|\mathbf h\right\|$ is the Euclidean norm of $\mathbf h$; $\lceil \cdot\rceil$ and $\lfloor \cdot\rfloor$ denote the integer ceiling and integer floor, respectively; $[\cdot]_{:,\mathcal{B}_k}$ and $[\cdot]_{\mathcal{B}_k,:}$ represent the
sub-matrices by collecting the related $\mathcal{B}_k$ columns or rows, respectively.

\section{System and Channel Model}\label{sec:2}
\begin{figure}[t]
      \centering
     \includegraphics[width=90mm]{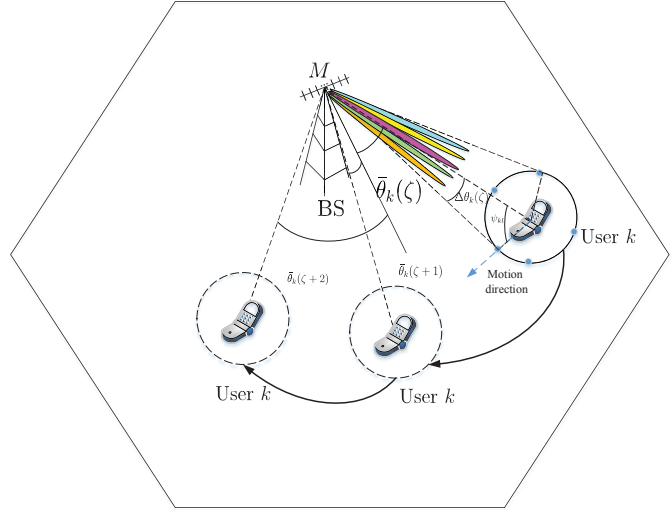}
     \caption{ System model. Users in the located cell are surrounded by a circle of local scatterers and
     the central DOA and AS of user-$k$ are $\bar\theta_k(\zeta)$ and $\Delta\theta_k(\zeta)$, respectively.}
     \label{fig:systemmodel}
\end{figure}

Consider a multiuser massive MIMO system shown in Fig. \ref{fig:systemmodel}, where base station (BS) is equipped with $M$ antennas in the form of uniform linear array (ULA), and $K$ users with single antenna are randomly distributed in the cell.  The baseband channels between users and BS are assumed to be time-selective flat-fading in each block $\zeta$ that contains $N$ information symbols, and the corresponding $M\times 1$ uplink channel between user $k$ and BS at time slot $n$ of block $\zeta$ can be expressed as \cite{globalcom,jinshi,GiannakisBEM}
\begin{equation}\label{equ:channelmodel}
\mathbf h_{k}(n)=\frac{1}{\sqrt{P}}\sum_{p=1}^P \alpha_{kp}e^{-j(2\pi f_d nT_s\cos\varphi_{kp}+\phi_{kp})}\mathbf a[\theta_{kp}(\zeta)],\ 0\leq n\leq N-1,
\end{equation}
where $P$ is the number of the multi-incoming rays, $\alpha_{kp}$ is the complex gain of the $p$-th ray, $f_d$ and $T_s$ are the maximum Doppler frequency and the system sampling period respectively, $\varphi_{kp}$ is the angle between uplink transmitted signal and the motion direction of user-$k$, and $\phi_{kp}$ represents  the initial phase, which is uniformly distributed in $[0,2\pi]$. Moreover, $\mathbf a[\theta_{kp}(\zeta)]\in\mathbb{C}^{M\times 1}$ is the steering vector and its $m$-th element can be expressed as
\begin{equation}\label{equ:steeringvector}
\{\mathbf a[\theta_{kp}(\zeta)]\}_m=e^{j\frac{2\pi md}{\lambda}\sin\theta_{kp}(\zeta)},m=0,\dots,M-1,
\end{equation}
where $d$ represents the antenna spacing, $\lambda$ is the signal carrier wavelength, and $\theta_{kp}(\zeta)$ represents DOA of the $p$-th ray of user $k$ seen by BS in block $\zeta$.
Denote the central DOA of user $k$ in block $\zeta$ as $\bar{\theta}_{k}(\zeta)$, and then each ray of user $k$ can be expressed as
\begin{equation}\label{equ:anglespread}
\theta_{kp}(\zeta)=\bar{\theta}_{k}(\zeta)+\tilde{\theta}_{kp}(\zeta),
\end{equation}
where $\tilde{\theta}_{kp}(\zeta)$ is the corresponding random AS satisfying $|\tilde{\theta}_{kp}(\zeta)|\leq\Delta \theta_k(\zeta)$, and $\Delta \theta_k(\zeta)$ is the maximum AS of user $k$, which is small for massive MIMO systems.

Moreover, $\tilde{\theta}_{kp}(\zeta)$ could be assumed as independent and identically distributed random variables satisfying \cite{2D-ESPRIT}
\begin{equation}\label{equ:spreadindependence}
\mathbb{E}\{\tilde{\theta}_{kp}(\zeta)\tilde{\theta}_{\tilde{k}\tilde{p}}\tilde{(\zeta)}\}=\sigma_k^2\delta(k-\tilde{k})\delta(p-\tilde{p})\delta(\zeta-\tilde{\zeta}),
\end{equation}
where $\sigma_k^2$ is the variance of $\tilde{\theta}_{kp}(\zeta)$. Since the spatial location of the users changes on the order of seconds, the DOA information of users seen by the BS can be viewed as unchanged in each data block $\zeta$, but  might change from block to block.


\section{Problem Formulation} \label{sec:3}
For array with large number of antennas, channel vector would exhibit many new features in both spatial domain and temporal  domain, which can be used for channel tracking.
\subsection{Spatial Domain Channel Representation with SBEM}
Let us define the $M$-point DFT of the channel vector $\tilde{\mathbf{h}}_k(n)$ as
\begin{equation}\label{equ:DFT}
\tilde{\mathbf h}_k(n)=\mathbf{F}\mathbf{h}_k(n), \ \ \ \ \ \ \textup{for}\  k=1,\ldots,K,
\end{equation}
where $\mathbf F$ is the $M\times M$ DFT matrix with $(p,q)$-th element $\left[\mathbf F\right]_{pq}=\frac{1}{\sqrt{M}}e^{-j\frac{2\pi}{M}pq}$.

According to (\ref{equ:DFT}), the $q$-th component of $\tilde{\mathbf h}_k(n)$ can be computed as
\begin{align}\label{equ:channeldftbjk}
\left[\tilde{\mathbf h}_k(n) \right]_{q}&=
\frac{1}{\sqrt{MP}}\sum_{p=1}^{P}\sum_{m=0}^{M-1}\alpha_{kp}e^{\vartheta_{kp}}e^{-j\left[\frac{2\pi}{M}mq-\frac{2\pi}{\lambda}md \sin\theta_{kp}(\zeta)\right]}\notag\\
&=\frac{1}{\sqrt{MP}}\sum_{p=1}^{P}\alpha_{kp}e^{(\vartheta_{kp}-j\frac{M-1}{2}\eta_{kp})}
\frac{\sin(\frac{M\eta_{kp}}{2})}
{\sin(\frac{\eta_{kp}}{2})},
\end{align}
where $\vartheta_{kp}=-j(2\pi f_d nT_s\cos\varphi_{kp}+\phi_{kp})$ and $\eta_{kp}=\frac{2\pi}{M}q-\frac{2\pi d}{\lambda}\sin\theta_{kp}(\zeta)$.

Due to both the narrow AS and the large number of antennas, $\tilde{\mathbf h}_k(n)$ is a highly sparse vector and most power is contained in a small set, which is called as  \emph{spatial signature} index (SSI) set and is denoted as $\mathcal{B}_k(\zeta)$. Besides, the left bound of $\mathcal{B}_k(\zeta)$ is determined by the leftmost ray $\bar\theta_k(\zeta){-}\Delta\theta_k(\zeta)$, while the right bound of $\mathcal{B}_k(\zeta)$ is determined by the rightmost ray $\bar\theta_k(\zeta){+}\Delta\theta_k(\zeta)$. When $M\rightarrow \infty$, the size of $\mathcal{B}_k(\zeta)$ is determined by the central DOA $\bar\theta_k(\zeta)$ and AS $\Delta\theta_k(\zeta)$\cite{xie} as
\begin{align}\label{equ:multi-rayrange}
     B_k= \left |\mathcal{B}_k(\zeta)\right|&{\approx} \left\lceil M\frac{d}{\lambda}\sin[\bar\theta_k(\zeta){+}\Delta\theta_k(\zeta)] \right\rceil
      {-}\left\lfloor M\frac{d}{\lambda}\sin[\bar\theta_k(\zeta){-}\Delta\theta_k(\zeta)] \right\rfloor \notag \\
       &{\approx}\left \lceil2M\frac{d}{\lambda}\left|\cos\bar\theta_k(\zeta)\right|\Delta\theta_k(\zeta)\right\rceil,
\end{align}
Moreover, the central index of  $\mathcal B_k(\zeta)$, denoted as  $q_{c,k}(\zeta)$,  is determined purely by the central DOA $\bar{\theta}_k(\zeta)$ as
\begin{equation}\label{equ:q_theta}
\begin{aligned}
{q_{c,k}(\zeta)}\approx\left\lfloor\frac{M d}{\lambda}\sin {\bar\theta}_k(\zeta)\right\rceil.
\end{aligned}
\end{equation}

\begin{figure}[t]
\centering
\includegraphics[width=90mm]{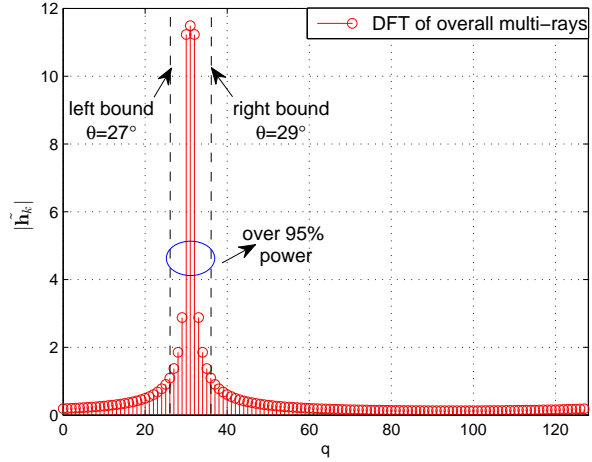}
\vspace{-2em}
\caption{Example of the DFT of channel model $\mathbf h_k(n)$ with DOA inside $[27^\circ,29^\circ]$ and $M=128, d=\lambda/2$.
 Since the AS is unchanged for the whole interval, the position of $\mathcal{B}_k$ is also the same for
 all $0\leq n\leq N-1$. \label{fig:multi-raypower}}
\end{figure}

An example of with AS $[27^\circ,29^\circ]$ is given in Fig. \ref{fig:multi-raypower} for $M=128$, where the DFT of entire multi-rays is illustrated. The SSI  $\mathcal{B}_k(\zeta)$ containing $\eta=95\%$ of the total power is $11$, which is very small in comparison with  $M$.

Therefore,  the channel vector $\mathbf{h}_k(n)$ could be expanded by a limited number of orthogonal basis as
\begin{align}\label{equ:DFTBEM}
\mathbf h_{k}(n)&{=}\mathbf F^H\tilde{\mathbf h}_{k}(n)
  {\approx}\left[\mathbf F^H\right]_{:,\mathcal{B}_{k}(\zeta)}\!\!\left[\tilde{\mathbf h}_{k}(n)\right]_{\mathcal{B}_{k}(\zeta),:}
    =\sum_{q\in\mathcal{B}_{k}(\zeta)} \tilde{h}_{k,q}(n) \mathbf{f}_q,
\end{align}
where $\tilde{h}_{k,q}(n)\triangleq [\tilde{\mathbf h}_k(n)]_q$,  and $\mathbf f_q$
is the orthogonal basis coming from the $q$-th column of $\mathbf F^H$.  Equation (\ref{equ:DFTBEM}) is named as spatial basis expansion model (SBEM) \cite{xie}.

\subsection{Time Domain Channel Representation with CE-BEM}
\begin{property}\label{property:2}
The components of $\tilde{\mathbf h}_k(n)$, $\tilde{h}_{k,q}(n), q=0,{\ldots},M{-}1$ are band-limited, and the maximum bandwidth of the power spectra is equal to the maximum Doppler frequency $f_d$ for massive MIMO system.
\end{property}

\begin{proof}
Let us define the time domain discrete correlation matrix of $\mathbf h_k(n)$ as $\mathbf R_k(m)\triangleq \mathbb E\{\mathbf h_k(n)\mathbf h_k^H(n+m)\}$, and the $(i,l)$-th component of $\mathbf R_k(m)$ is given as
\begin{align} \label{equ:correlation}
    [\mathbf R_k(m)]_{i,l}&=\mathbb E\{h_{k,i}(n)h^*_{k,l}(n+m)\}\notag\\
    &=\frac{1}{P}\sum_{p=1}^P \mathbb E\{|\alpha_{kp}|^2\}\mathbb E\{e^{-j2\pi f_d mT_s \cos\varphi_{kp}}\}\cdot \mathbb E\{e^{j\frac{2\pi d}{\lambda}(i-l)\sin\theta_{kp}}\}.
\end{align}

The proof process is different from the conventional Clarke's reference model \cite{chengshanxiao}, where the incoming angles of user seen by BS are assumed to be uniformly distributed in $[0,2\pi]$, the AS of user $k$ keeps within a small range, i.e., $[\theta_k(\zeta)-\Delta\theta_k(\zeta),\theta_k(\zeta)+\Delta\theta_k(\zeta)]$, while $\varphi_{kp}$ is assumed to be randomly distributed at $[0,2\pi]$ for the random mobility of users. Consequently, the expectation in \eqref{equ:correlation} is mainly focused on $\varphi_{kp}$, and then it holds that
\begin{align}\label{equ:timecorrelation}
    [\mathbf R_k(m)]_{i,l}&=J_0(2\pi f_d m T_s)\cdot g(2\pi d/\lambda(l-i)),
\end{align}
where $J_0(x)$ is the first kind zero-order Bessel functions, which is given by
\begin{align} \label{equ:j0}
J_0(x)=\frac{1}{2\pi}\int_{-\pi}^\pi e^{-jx\cos y}dy,
\end{align}
while $g(x)$ is defined as
\begin{align} \label{equ:j1}
g(x)=\frac{1}{2\Delta\theta_k}\int_{\theta_k-\Delta\theta_k}^{\theta_k+\Delta\theta_k}e^{-jx\sin y}dy.
\end{align}
Note that $g(x)$ does not affect the bandwidth of \eqref{equ:timecorrelation}. Besides, according to \cite{Clarke}, the power spectrum of $J_0(2\pi f_d m T_s)$ is the ``U-shape'' function,
  \begin{align}\label{equ:J_0}
       S_{J_0}(f)=\frac{1}{\pi f_d\sqrt{1-f^2/f_d^2}},\quad f\in[-f_d,f_d],
  \end{align}
and hence the bandwidth of \eqref{equ:timecorrelation} is a constant $f_d$. Interestingly,  the range of the AS (spatial property) will not affect time-domain bandwidth of ${h}_{k,q}(n)$ (frequency property).

Since $\tilde{h}_{k,q}(n)=\sum_{i=0}^{M-1}h_{k,i}(n)e^{-j\frac{2\pi}{M}iq}$, we have
\begin{align}
\mathbb E\{\tilde{h}_{k,q}(n)[\tilde{h}_{k,q}(n+m)]^*\}
=\sum_{i=0}^{M-1}\sum_{l=0}^{M-1}\mathbb E\{h_{k,i}(n)[h_{k,l}(n+m)]^*\}e^{-j\frac{2\pi }{M}q(i-l)},
\end{align}
where $\mathbb E\{\tilde{h}_{k,q}(n)[\tilde{h}_{k,q}(n+m)]^*\}$ is the superposition of multiple band-limited signals with the same bandwidth $f_d$. Therefore, the components $\tilde{h}_{k,q}(n), q=0,{\ldots},M{-}1$ are all band-limited and upper bounded by $f_d$.
\end{proof}

From Property \ref{property:2}, we know $ \tilde{h}_{k,q}(n)$ could be \emph{timely} expanded by $\mu\ll M$ limited orthogonal time basis to capture the rapid variation of $\tilde {h}_{k,q}(n)$ for massive MIMO systems \cite{gao_relay} as
\begin{align}\label{equ:TDBEM}
  \tilde{h}_{k,q}(n)=\sum_{r=0}^{\mu}\gamma_{k,q}^re^{j2\pi(r-\mu/2)n/N}, \ \ 0\leq n\leq N-1,
\end{align}
where $\gamma_{k,q}^r$'s are the invariant coefficients, while the order $\mu$ is a function of the
channel bandwidth $f_d$ and the length of sampling interval $NT_s$. Equation (\ref{equ:TDBEM}) is also known as the complex exponential basis expansion model (CE-BEM).
\subsection{Channel  Dimension Reduction with ST-BEM}\label{sec:ST-BEM}
\begin{figure}[t]
\centering
\includegraphics[width=90mm]{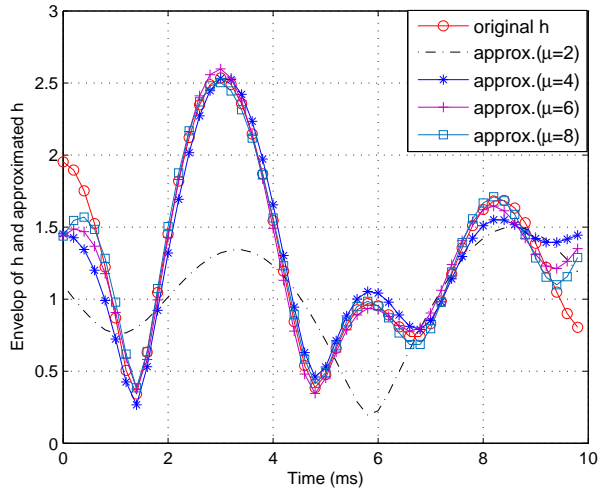}
\vspace{-1em}
\caption{CE-BEM approximation of $\tilde{h}_{k,2}(n), n=1,\ldots,N$ with different values of $\mu$.
\label{fig:BEMapproximate}}
\end{figure}

According to previous discussions, the overall time-varying channel vector $\mathbf{h}_k(n)$ can be jointly expanded as
\begin{align}\label{equ:equal}
  \mathbf h_k(n) &\approx \sum_{q\in\mathcal{B}_k(\zeta)} \tilde{h}_{k,q}(n)\mathbf f_q=
  \sum_{q\in\mathcal{B}_k(\zeta)} \sum_{r=0}^\mu \gamma_{k,q}^r e^{j2\pi(r-\mu/2)n/N}\mathbf f_q\notag\\
  &=\sum_{q\in\mathcal{B}_k(\zeta)} \bm{\gamma}_{k,q}^T \mathbf c_n \mathbf f_q, \ \ \  n=0,\ldots,N-1,
\end{align}
where $\bm \gamma_{k,q}{=}[\gamma_{k,q}^0,\ldots,\gamma_{k,q}^\mu]^T$ and
$\mathbf c_n{=}[e^{-j\frac{2\pi n}{N}\frac{\mu}{2}},\ldots,e^{j\frac{2\pi n}{N}\frac{\mu}{2}}]^T$. Moreover, (\ref{equ:equal}) can be named as spatial-temporal basis expansion model (\emph{ST-BEM}), since it represents the channel as an combination of the SBEM and CE-BEM.

It is worth noticing that $\mu$ should satisfy  $\mu \geq 2\lceil f_dNT_s\rceil$ to offer sufficient degrees of freedom \cite{temporalBEM1}. In order to provide an explicit illustration, an example of $\tilde{h}_{k,2}(n)$ is offered in Fig.~\ref{fig:BEMapproximate}, where the simulation parameters are chosen as $M=128,d=\lambda/2, f_d=200$Hz, $T_s=0.1$ ms and $N=100$. It can be found that when $\mu\geq 2\lceil f_dNT_s\rceil=4$, the approximation of $\tilde{h}_{k,2}(n)$ is well, while the ambiguous estimation happens for $\mu=2$ due to the lack of the sufficient sampling degrees of freedom.

\section{Tracking the Spatial Information with Parameter Learning} \label{sec:4}
In this section, we provide a blind uplink spatial information tracking method based on the uplink received symbols. With (\ref{equ:equal}), channel tracking problem is transmitted to tracking SSI set $\mathcal B_k(\zeta)$ and estimating the CE-BEM coefficients $\bm \gamma_{k,q}$.
Furthermore, since $\mathcal B_k(\zeta)$ is determined by central DOA  $\bar{\theta}_{k}(\zeta)$ and AS $\Delta \theta_k(\zeta)$ according to (\ref{equ:multi-rayrange}) and (\ref{equ:q_theta}), tracking the spatial information can be decomposed into tracking $\bar{\theta}_{k}(\zeta)$ and $\Delta \theta_k(\zeta)$ separately.
\subsection{Central DOA $\bar{\theta}_k(\zeta)$ Modelling }
According to \cite{xie}, users that can be simultaneously scheduled should have distinct spatial information to decrease user interference. Therefore, users can be allocated into the same group if their SSI sets do not overlap and also guarded by a interval $\Omega$, i.e.,
\begin{align}
\mathcal{B}_k(\zeta)\cap\mathcal{B}_l(\zeta)=\emptyset,\ \ \tu{dist}\left[\mathcal{B}_k(\zeta),\mathcal{B}_l(\zeta)\right]\geq \Omega,
\end{align}
Note that the users are scheduled according to their DOA information, and hence the corresponding transmission scheme can be named as angle division multiple access (ADMA). One can find much more detailed descriptions and methodologies
of ADMA in \cite{xie,xieCR,overview}. The initial DOA information of each user can be derived from method in  \cite{xie}. Interested readers could refer to \cite{xie} for more details.

Assuming that all $K$ users are divided into $G$ groups, and the $g$-th group is denoted as $\mathcal{U}_g$ with $K_g$ users. When users in group $\mathcal{U}_g$ are scheduled, the received signals at BS in time slot $n$ of block $\zeta$ can be expressed as
\begin{equation}\label{equ:received}
\mathbf x(n)=\sum_{k=1}^{K_g}\mathbf h_k(n)s_k(n)+\mathbf w(n)=\sum_{k=1}^{K_g}\frac{ s_k(n)}{\sqrt{P}}\sum_{p=1}^P \alpha_{kp}e^{\xi_{kp}}\mathbf a[\theta_{kp}(\zeta)]+\mathbf w(n),
\end{equation}
where $s_k(n)$ denotes the transmitted signal of user $k$, and $\mathbf w(n)$ is the noise vector whose elements are i.i.d. with distribution $\mathcal{CN}(0,\sigma_n^2)$.

Since the data transmission also possesses the  Vandermonde structure of the channel \cite{data-aided}, the DFT of $\tilde{\mathbf x}(n)$ can still tell the spatial signature of the instantaneous $\mathbf h_k(n)$. According to (\ref{equ:channeldftbjk}), it holds that
\begin{equation}\label{xk}
[\tilde{\mathbf x}(n)]_{\mathcal B_k(\zeta)}=\left[s_k(n)\mathbf F\mathbf h_{k}(n)\right]_{\mathcal B_k(\zeta)}.
\end{equation}

Note that $[\tilde{\mathbf x}(n)]_{\mathcal B_k(\zeta)}$ serves a humble observation of the spatial signature. Then a straightforward way to track the SSI set $\mathcal{B}_k(\zeta)$ is to obtain the size of $\mathcal{B}_k(\zeta)$ from the size of $[\tilde{\mathbf x}(n)]_{\mathcal B_k(\zeta)}$, while to derive central SSI $q_{c,k}(\zeta)$ of $\mathcal{B}_k(\zeta)$ from the central SSI of $[\tilde{\mathbf x}(n)]_{\mathcal B_k(\zeta)}$.  The corresponding method can be named as \emph{DFT searching}. However, when $M$ is not infinite, power leakage effect \cite{xie} would actually distort the central position as well as the size of $\mathcal{B}_k(\zeta)$. Therefore, we will propose a more effective way to track the spatial information.

From (\ref{xk}), we can derive the observed central SSI $q_{c,k}(\zeta)$, which provides us a practical measurement of central DOA  $\bar{\theta}_k(\zeta)$. According to \eqref{equ:q_theta}, the relationship between $q_{c,k}(\zeta)$ and $\bar{\theta}_{k}(\zeta)$ can be represented by
\begin{equation}\label{equ:qc}
\begin{aligned}
q_{c,k}(\zeta) =\frac{dM}{\lambda}\sin {{\bar\theta}_k(\zeta)}+u_k(\zeta),
\end{aligned}
\end{equation}
where $u_k(\zeta)$ is the measurement noise meeting i.i.d. Gaussian distribution $\mathcal{CN}(0,Q_{u_k})$. Equation~\eqref{equ:qc} can be viewed as the \emph{measurement equation}.

Besides, the one-order Markov process \cite{mmdoa,liulingjia0} could be utilized to describe the variation of central DOAs as
\begin{equation}\label{equ:kinematic}
\begin{aligned}
\bar{\theta}_k(\zeta) =\bar{\theta}_k(\zeta-1)+\omega_k(\zeta-1),
\end{aligned}
\end{equation}
where $\bar{\theta}_k(\zeta-1)$ are the central DOA in block $\zeta-1$. Moreover, $\omega_k(\zeta-1)$ is the system noise, which meets $\mathbb{E}[\omega_k(\zeta-1)^H\omega_k(\zeta-1)]={Q}_{\omega_k}$. Equation (\ref{equ:kinematic}) can be viewed as the\emph{ system equation}.

\subsection{EM based Parameter Learning}
Before proceeding, we need first derive the unknown parameters ${Q}_{ \omega_k}$ and $Q_{u_k}$. Denote  $\boldsymbol\beta_k=[ Q_{\omega_k},Q_{u_k}]$ and $\mathbf q_k=[q_{c,k}(0),q_{c,k}(1),\ldots,q_{c,k}(\varsigma-1)]^T$ as the unknown parameter vector and the observation vector, respectively, where $\varsigma$ is the dimension of the observation vector for parameter learning. A data-aided maximum likelihood (DA-ML) estimator for $\boldsymbol\beta_k$ can be formulated as
\begin{align}
\boldsymbol{\hat\beta}_k=&\max p(\mathbf q_k|\boldsymbol\beta_k).
\end{align}
Nevertheless, the direct ML estimation of $\boldsymbol\beta_k$ is not feasible for its large computation overhead. An effective solution is to search for the ML solution iteratively via the EM algorithm, which consists of two steps, namely \emph{the expectation step} and \emph{the maximization step}. Besides, it has been verified that $\boldsymbol{\hat\beta_k}^{(l)}$ can converge to one stationary point of the likelihood function $p(\mathcal R|\boldsymbol{\beta_k})$ under fairly general conditions \cite{EM}.

Let us denote $\bar{\boldsymbol \theta}_k=[\bar{ \theta}_k(0),\dots,\bar{ \theta}_k(\varsigma-1)]$. Using the EM algorithm, the vector $\mathbf q_k$ forms
the incomplete data set $\mathcal R$, while the vectors  $\mathbf q_k$ together with $\bar{\boldsymbol \theta}_k$ form
the complete data set $\mathcal Z$. During the $l$-th iteration, we firstly compute the objective function $\boldsymbol L({\boldsymbol\beta_k},\hat{\boldsymbol\beta_k}^{(l-1)})$ in the expectation step as
\begin{align}
\boldsymbol L({\boldsymbol\beta_k},\hat{\boldsymbol\beta_k}^{(l-1)})=\int_{\mathcal Z}p(\mathcal Z|\mathcal R,\hat{\boldsymbol \beta_k}^{(l-1)})\ln p(\mathcal Z|\boldsymbol{\beta_k}) d\mathcal Z.
\end{align}

Note that $\boldsymbol L({\boldsymbol\beta_k},\hat{\boldsymbol\beta_k}^{(l-1)})$ depends on the estimation $\hat{\boldsymbol\beta}^{(l-1)}$ in the $(l-1)$th iteration and the trial value ${\boldsymbol\beta_k}$. Since the set $\mathcal R$ is known, the function $\boldsymbol L({\boldsymbol\beta_k},\hat{\boldsymbol\beta_k}^{{(l-1)}})$  can be
rewritten as
\begin{align}\label{op}
\boldsymbol L({\boldsymbol\beta_k},\hat{\boldsymbol\beta_k}^{(l-1)})=&\int_{{\bar{\boldsymbol \theta}_k}}
p(\bar{\boldsymbol \theta}_k|\mathcal R,\boldsymbol{\hat\beta_k}^{(l-1)})\ln\left[p(\mathcal R|\bar{\boldsymbol \theta}_k,\boldsymbol{ \beta_k})p(\bar{\boldsymbol \theta}_k|\boldsymbol{\beta_k})\right]d{\bar{\boldsymbol \theta}_k}\notag\\
=&\int_{\bar{\boldsymbol \theta}_k}
p(\bar{\boldsymbol \theta}_k|\mathcal R,\boldsymbol{\hat\beta_k}^{(l-1)})\left[
\ln p(\mathcal R|\bar{\boldsymbol \theta}_k,\boldsymbol{\beta_k})
+\ln p(\bar{\boldsymbol \theta}_k|\boldsymbol{\beta_k})\right]d{\bar{\boldsymbol \theta}_k}\notag\\
=&\mathbb E\left\{\ln p(\mathcal R|\bar{\boldsymbol \theta}_k,\boldsymbol{\beta_k})|
\mathcal R,\boldsymbol{\hat\beta_k}^{(l-1)}\right\}+
\mathbb E\left\{\ln p(\bar{\boldsymbol \theta}_k|\boldsymbol{\beta_k})|
\mathcal R,\boldsymbol{\hat\beta_k}^{(l-1)}\right\}.
\end{align}

According to the system equation \eqref{equ:kinematic} and measurement equation \eqref{equ:qc}, we have

\begin{align}
\ln p(\mathbf q_k|\bar{\boldsymbol \theta}_k,\boldsymbol{\beta_k})&=\sum_{\zeta=0}^{\varsigma-1}\ln p(\mathbf {q_k}_\zeta|\bar{\boldsymbol \theta}_{k\zeta},\boldsymbol\beta_k)\notag \\
&=\sum_{\zeta=0}^{\varsigma-1}\ln \frac{1}{\sqrt{2\pi Q_{u_k}}}\exp\frac{-\left(q_{c,k}(\zeta) -\frac{dM}{\lambda}\sin {{\bar\theta}_k(\zeta)}\right)^2}{{2 Q_{u_k}}},\\
\ln p(\bar{\boldsymbol \theta}_k|\boldsymbol\beta_k)&=-\frac{1}{2}\varsigma\ln 2\pi Q_{\omega_k}-\sum_{\zeta=0}^{\varsigma-1}\frac{\left(\bar\theta_{k}(\zeta) -\bar\theta_{k}(\zeta-1)\right)^2}{2 Q_{\omega_k}}.
\end{align}
Therefore, it can be further derived that
\begin{align}
&\mathbb E\left\{\ln p(\mathbf{q}_k|\boldsymbol{\bar\theta}_k,\boldsymbol\beta_k)|\mathbf q_k,\boldsymbol{\hat\beta}_k^{(l-1)}\right\}=-\frac{1}{2}\varsigma \ln 2\pi Q_{u_k}^{(l-1)}-\frac{\sum_{\zeta=0}^{\varsigma-1} \left[q_{c,k}(\zeta)\right]^2}{2Q_{u_k}^{(l-1)}}+\notag \\
&\frac{\sum_{\zeta=0}^{\varsigma-1} \left[q_{c,k}(\zeta)\right]\frac{dM}{\lambda}\mathbb E\left\{ \sin {{\bar\theta}_k(\zeta)}| \mathbf q_k,\boldsymbol{\hat\beta}_k^{(l-1)} \right\}}{Q_{u_k}^{(l-1)}}\!-\!\frac{\sum_{\zeta=0}^{\varsigma-1} \left[\frac{dM}{\lambda}\right]^2E\left\{ \sin {{\bar\theta}_k^2(\zeta)}| \mathbf q_k,\boldsymbol{\hat\beta}_k^{(l-1)} \right\}}{2Q_{u_k}^{(l-1)}},\\
&\mathbb E\left\{p(\bar{\boldsymbol \theta}_k|\boldsymbol\beta_k)|\mathbf q_k,\boldsymbol{\hat\beta}_k^{(l-1)}\right\}
=-\frac{1}{2}\varsigma\ln 2\pi Q_{\omega_k}^{(l-1)}-\frac{\sum_{\zeta=0}^{\varsigma-1}\mathbb E\left\{{{\bar\theta}_k^2(\zeta)}| \mathbf q_k,\boldsymbol{\hat\beta}_k^{(l-1)} \right\}}{2 Q_{\omega_k}^{(l-1)}}\notag \\&-\frac{\sum_{\zeta=0}^{\varsigma-1}\mathbb E\left\{{{\bar\theta}_k^2(\zeta-1)}| \mathbf q_k,\boldsymbol{\hat\beta}_k^{(l-1)} \right\}}{2 Q_{\omega_k}^{(l-1)}}+\sum_{\zeta=0}^{\varsigma-1}\frac{\mathbb E\left\{{{\bar\theta}_k(\zeta)}{{\bar\theta}_k(\zeta-1)}| \mathbf q_k,\boldsymbol{\hat\beta}_k^{(l-1)} \right\}}{Q_{\omega_k}^{(l-1)}}.
\end{align}

Secondly, we turn to the maximization of $\boldsymbol L({\boldsymbol\beta_k},\hat{\boldsymbol\beta_k}^{(l-1)})$ as
\begin{align}
\boldsymbol{\hat\beta_k}^{(l)}=\arg\max_{{\boldsymbol\beta_k}}\left\{\boldsymbol L({\boldsymbol\beta_k},\hat{\boldsymbol\beta_k}^{(l-1)})\right\}.
\end{align}

 The solution of (\ref{op}) can be derived from $\frac{\partial L\left({\boldsymbol\beta_k},\hat{\boldsymbol\beta_k}^{(l-1)}\right)}{\partial Q_{\omega_k}}=0$ and $\frac{\partial L\left({\boldsymbol\beta_k},\hat{\boldsymbol\beta_k}^{(l-1)}\right)}{\partial Q_{u_k}}=0$ as
\begin{align}\label{result1}
Q_{u_k}=&\frac{1}{\varsigma}\sum_{\zeta=0}^{\varsigma-1} \left[q_{c,k}(\zeta)\right]^2+\frac{1}{\varsigma}\sum_{\zeta=0}^{\varsigma-1} \left[\frac{dM}{\lambda}\right]^2E\left\{ \sin {{\bar\theta}_k^2(\zeta)}| \mathbf q_k,\boldsymbol{\hat\beta}_k^{(l-1)} \right\}\notag\\
&-\frac{2}{\varsigma}\sum_{\zeta=0}^{\varsigma-1}\frac{q_{c,k}(\zeta)dM}{\lambda}\mathbb E\left\{ \sin {{\bar\theta}_k(\zeta)}| \mathbf q_k,\boldsymbol{\hat\beta}_k^{(l-1)} \right\},\\
\label{result2}Q_{\omega_k}=&\frac{1}{\varsigma}\sum_{\zeta=0}^{\varsigma-1}\mathbb E\left\{{{\bar\theta}_k^2(\zeta)}| \mathbf q_k,\boldsymbol{\hat\beta}_k^{(l-1)} \right\}+\frac{1}{\varsigma}\sum_{\zeta=0}^{\varsigma-1}\mathbb E\left\{{{\bar\theta}_k^2(\zeta-1)}| \mathbf q_k,\boldsymbol{\hat\beta}_k^{(l-1)} \right\}\notag\\
&-\frac{2}{\varsigma}\sum_{\zeta=0}^{\varsigma-1}\mathbb E\left\{{{\bar\theta}_k(\zeta)}{{\bar\theta}_k(\zeta-1)}| \mathbf q_k,\boldsymbol{\hat\beta}_k^{(l-1)} \right\}.
\end{align}

When we can derive the statistics of the system noise and measurement noise in~(\ref{result1})~and~(\ref{result2}), the central DOA can be tracked by recursive predicting and updating with~UKF.

\subsection{Tracking Central DOA $\bar{\theta}_k(\zeta)$ with UKF and URTSS}
Based on Bayesian filtering and smoothing framework \cite{EM}, the tracking of the central DOA can then be partitioned into two phases: the forward tracking and the backward smoothing. During the former phase, Bayesian filter is adopted to sequentially estimate the current state, while during the latter phase, the corresponding Bayesian smoother is utilized to reconstruct the system states. Since the observation vector $\mathbf q_k$ is not linear with respect to the central DOA vector $\bar {\boldsymbol \theta}_{k}$, the dynamic state space is not a linear Gaussian state space either. Thus,
the conventional Kalman filter (KF) and Rauch-Tung-Striebel smoother (RTSS) can not be utilized. Though extended Kalman filter (EKF) is a widely adopted nonlinear filter but cannot be used here since Taylor series expansion of EKF has large truncation error, which would severely degrade the channel tracking performance. In the following, we would utilize unscented Kalman filter (UKF) to implement forward tracking and unscented Rauch-Tung-Striebel smoother (URTSS) to achieve the backward smoothing respectively for each iteration~ $l$~\cite{UKF,URTSS}.

Let us  define the following posteriori statistics to describe the distribution of the system states:\begin{align}
\bar{\theta}_{k}^{s,(l)}(\zeta)=&\mathbb E\left\{\bar {\theta}_k(\zeta)|\mathbf q_k,\boldsymbol{\beta}_k^{(l-1)}\right\},\\
P_{k}^{s,(l)}(\zeta)=&\mathbb E\left\{\left[\bar\theta_{k}(\zeta)-\mathbf{{\bar\theta}}_{k}^{s,(l)}(\zeta)\right]\left[\bar\theta_{k}(\zeta)-\mathbf{{\bar\theta}}_{k}^{s,(l)}(\zeta)\right]^H|\mathbf q_k,\boldsymbol{\beta}_k^{(l-1)}\right\},\\ C_{k,\zeta-1,\zeta}^{s,(l)}=&
\mathbb E\left\{\left[\bar\theta_{k}(\zeta-1)-\mathbf{{\bar\theta}}_{k}^{s,(l)}(\zeta-1)\right]\left[\bar\theta_{k}(\zeta)-\mathbf{{\bar\theta}}_{k}^{s,(l)}(\zeta)\right]^H|\mathbf q_k,\boldsymbol{\beta}_k^{(l-1)}\right\}.
\end{align}

Besides, it can be further derived that
\begin{align}
\mathbb E\left\{\bar\theta_{k}(\zeta)\bar\theta_{k}^H(\zeta)|\mathbf q_k,\boldsymbol{\beta}_k^{(l-1)}\right\}
=&{\bar\theta}_{k}^{s,(l)}(\zeta)[{\bar\theta}_{k}^{s,(l)}(\zeta)]^H+
 P_{k}^{s,(l)}(\zeta)
,\\
\mathbb E\left\{\bar\theta_{k}(\zeta-1)\bar\theta_{k}^H(\zeta)|\mathbf q_k,\boldsymbol{\beta}_k^{(l-1)}\right\}
=&{\bar\theta}_{k}^{s,(l)}(\zeta-1)[{\bar\theta}_{k}^{s,(l)}(\zeta)]^H+
 C_{k,\zeta-1,\zeta}^{s,(l)}
.
\end{align}

Now, let us start estimating the posteriori statistics
$\bar \theta_{k}^{s,(l)}(\zeta)$, $P_{k}^{s,(l)}(\zeta)$, and $C_{k,\zeta-1,\zeta}^{s,(l)}$
conditioned on $\mathbf q_k$ and $\boldsymbol{\bar\theta}_k^{(l-1)}$.

\subsubsection{\textbf{Forward Tracking by UKF}}

 According to the procedure of UKF, a total of $2R+1$ sigma points $\chi_k^{(i)}$ are calculated from  $\sqrt{ P_k(\zeta-1)}$ and $ {\bar \theta}_k(\zeta-1)$ in block $\zeta-1$ to describe the central DOA distribution of each user for massive MIMO systems.
\begin{equation}\label{sigmapoints}
\begin{aligned}
\left\{\begin{array}{l}
\chi^{(0)}_k= {\bar \theta}_k(\zeta-1),i = 0,\\
\chi^{(i)}_k= {\bar \theta}_k(\zeta-1)+\left[\sqrt{R+\varepsilon} \sqrt{ P_k(\zeta-1)}\right],i=1,\dots,R,\\
\chi^{(i)}_k= {\bar \theta}_k(\zeta-1)-\left[\sqrt{R+\varepsilon} \sqrt{ P_k(\zeta-1)}\right],i=R+1,\dots,2R,
\end{array}\right.
\end{aligned}
\end{equation}
where $\varepsilon=\alpha^2(R+\kappa)-R$, $\alpha$ is a scaling parameter controlling the spread of sigma points around $ {\bar \theta}_k(\zeta-1)$, $R$ is the number of the system states, and $\kappa$ is a secondary scaling parameter, which is usually set as $3-R$.

Next, the system equation is applied to all the sigma points $\chi^{(i)}_k, i=0,\dots,2R$ to yield a cloud of the transformed posterior
points as
\begin{align}\label{equ:ukf5}
\iota_k^{(i)} = {\bf{f}}\left(\chi_k^{(i)}\right),
\end{align}
where $\bf{f}$ is the system state transformed function determined by (\ref{equ:kinematic}). Besides, the transformed points  $\iota_k^{(i)}$ can be viewed as predicted sigma points of the central DOA.

Then, the statistics of the posterior points $\iota_k^{(i)}$  are calculated to form the transformed mean $ {\bar \theta}_k^ -(\zeta)$  and covariance $P_k^-(\zeta)$ for central DOA tracking as
\begin{align}\label{equ:ukf6}
& {\bar \theta}_k^ -(\zeta)  = \sum_{i = 0}^{2R} W_m^{(i)}\iota_k^{(i)},\\
&P_k^-(\zeta) = \sum_{i=0}^{2R}W_c^{(i)} \left[\iota_k^{(i)}- {\bar \theta}_k^ -(\zeta)\right]\left[\iota_k^{(i)}- {\bar \theta}_k^ -(\zeta)\right]^H + Q_{\omega_k},
\end{align}
where $W_m^{(i)}$ and $ W_c^{(i)}$ are the weights for the mean and covariance respectively. Moreover, these weights can be derived as
\begin{align}\label{sigmapointsweight}
\left\{\begin{array}{l}
 W_m^{(0)}=\frac{\varepsilon}{R+\varepsilon},i = 0,\\
W_m^{(i)}=\frac{\varepsilon}{2(R+\varepsilon)},i=1,\dots,2R,
\\ W_c^{(0)}=\frac{\varepsilon}{R+\varepsilon}+1-\alpha^2+\beta,i = 0,\\
 W_c^{(i)}=\frac{\varepsilon}{2(R+\varepsilon)},i=1,\dots,2R,
\end{array}\right.
\end{align}
where $\beta$ is used to incorporate the prior knowledge of the distribution of the system states, and the optimal choice of $\beta$ for Gaussian distribution is normally taken as 2 \cite{UKF}.

Next, we turn to the prediction of the measurement states. Similarly, unscented transformation is used to approximate a probability distribution with the predicted ${\bar \theta}_k^-(\zeta)$ and $P_k^-(\zeta)$, and the predicted sigma points can be derived as
\begin{equation}\label{gettingsigmapointspredition}
\begin{aligned}
\left\{\begin{array}{l}
\chi^{(0)-}_k= {\bar \theta}_k^-(\zeta-1),i = 0,\\
\chi^{(i)-}_k= {\bar \theta}_k^-(\zeta-1)+\left[\sqrt{R+\varepsilon} \sqrt{ P_k^-(\zeta-1)}\right],i=1,\dots,R,\\
\chi^{(i)-}_k= {\bar \theta}_k^-(\zeta-1)-\left[\sqrt{R+\varepsilon} \sqrt{ P_k^-(\zeta-1)}\right],i=R+1,\dots,2R,
\end{array}\right.
\end{aligned}
\end{equation}
such that their mean and covariance are ${\bar \theta}_k^-(\zeta)$ and $P_k^-(\zeta)$ of the predicted central DOA.

Then, the measurement equation is applied to $\chi_k^{(i)}$ to form measurement sigma points $\xi_k^{(i)}$ as
\begin{equation}\label{u1}
\xi_k^{(i)} = \left\lceil\frac{dM}{\lambda}\sin \chi_k^{(i)-}\right\rfloor.
\end{equation}
Since $\xi_k^{(i)}$ represents the possible candidates of the predicted central SSI for the central DOA tracking, we can use the mean of $\xi_k^{(i)}$ to represent the predicted central SSI as
\begin{equation}\label{u4}
{{y}}_k^ -  = \sum\limits_{i = 0}^{2R+1} W_m^{(i)} \xi_k^{(i)}.
\end{equation}

Then, the forward Kalman gain for UKF can be derived as
\begin{equation}\label{equ:ukf8}
\begin{aligned}
{{{K}}_k} = {{{P}}_{xy}}{{P}}_{yy}^{ - 1},
\end{aligned}
\end{equation}
where $P_{xy}$ and $P_{yy}$ are the predicted state-measurement cross covariance and measurement covariance
respectively with the expressions:
\begin{align}\label{equ:u2}
&{{{P}}_{xy}} = \sum\limits_{i = 0}^{2R} {W_c^{(i)}} \left[\iota_k^{(i)}- {\bar \theta}_k^ -(\zeta)\right]\left[ \mathbf\xi_k^{(i)} -y_k^- \right]^T,\\
&{{{P}}_{yy}} = \sum\limits_{i = 0}^{2R} {W_c^{(i)}} \left[\mathbf\xi_k^{(i)} - {{y}}_k^ - \right]{\left( \mathbf\xi_k^{(i)}-y_k^-\right)^T} + Q_{u_k}.
\end{align}

At the last step of UKF, the filtered system states and the corresponding error covariance for central DOA tracking can be respectively obtained by
\begin{align}\label{equ:ukf9}
&\bar\theta_k(\zeta)= \bar \theta_k^-(\zeta)+K_k[q_{c,k}(\zeta) - y_k^- ],\\
&P_k(\zeta) =P_k^-(\zeta)-K_k P_{yy}K_k^T.
\end{align}

\begin{remark}
The complexity of the UKF based forward DOA tracking can be evaluated by the calculation of the $K_g \times K_g$ square roots of $P_k^-(\zeta)$, $P_k(\zeta)$ in (\ref{sigmapoints})-(\ref{gettingsigmapointspredition}) through the Cholesky factorization operation, whose complexity  is  $\mathcal{O}\left(\frac{K_g^3}{6}\right)$. In comparison, the covariance based channel tracking method needs to perform singular value decomposition, and the corresponding complexity is $\mathcal{O}\left({M^3}\right)$ ($K_g\ll M$). Therefore, the proposed  channel tracking method has a relatively lower complexity.
\end{remark}

\subsubsection{\textbf{Backward Smoothing with URTSS}}
 UKF can only derive the system states based on the previous measurement, but not the whole trajectory of the measurements. When we get the forward estimated values of $\bar\theta_k(\zeta)$ and $P_k(\zeta)$, we could use URTSS to obtain a more desirable estimation and the posteriori statistic of the central DOA  $\bar{\theta}_k^s(\zeta)$, $\zeta=\varsigma,\dots,0$.

The $2R+1$ sigma points $\chi^{s,(i)}_{k}$ can be derived from  the forward UKF as
\begin{equation}\label{sigmapoints2}
\begin{aligned}
\left\{\begin{array}{l}
\mathbf \chi^{s,(0)}_{k}=\bar{ \theta}_k(\zeta),i = 0,\\
\chi^{s,(i)}_{k}=\bar{ \theta}_k(\zeta)+\left[\sqrt{R+\varepsilon} \sqrt{ P_k(\zeta)}\right],i=1,\dots,R,\\
\chi^{s,(i)}_{k}=\bar{ \theta}_k(\zeta)-\left[\sqrt{R+\varepsilon} \sqrt{ P_k(\zeta)}\right],i=R+1,\dots,2R.
\end{array}\right.
\end{aligned}
\end{equation}

Then, the transformed sigma points $\iota_k^{s,(i)}$ and the predicted mean $y_{k}^ {s- }$ of the central DOA can be expressed as
\begin{align}\label{equ:ukf52}
&\iota_k^{s,(i)}= {\bf{f}}\left(\chi^{s,(i)}_{k}\right),\\
&y_{k}^ {s- } = \sum\limits_{i = 0}^{2R+1} W_m^{(i)}\iota_k^{s,(i)}.
\end{align}

Similar with the procedure of UKF \eqref{equ:ukf8}-\eqref{equ:u2}, the URTSS gain could be computed as
\begin{align}\label{equ:ukf82}
&K_{k}^s =C_{k,\zeta-1,\zeta}^{s}{P_{yy}^{s}}^{-1},
\end{align}
where the predicted cross covariance $C_{k,\zeta-1,\zeta}^{s}$ and the predicted covariance ${{P}}_{yy}^{s}$ can be separately derived as
\begin{align}\label{equ:ukf83}
&C_{k,\zeta-1,\zeta}^{s} = \sum\limits_{i = 0}^{2R} W_c^{(i)} \left[\chi^{s,(i)}_{k} -\bar{ \theta}_k(\zeta-1) \right]\left( \iota_k^{s,(i)} - y_\zeta^ {s- } \right)^T,\\
&{{P}}_{yy}^{s} = \sum\limits_{i = 0}^{2R} {W_c^{(i)}} \left( \iota_k^{s,(i)} - y_\zeta^ {s- } \right)\left( \iota_k^{s,(i)} - y_\zeta^ {s- } \right)^T + Q_{\omega_k}.
\end{align}

Afterwards, the smoothed central DOA and the corresponding error covariance of the central DOA can be acquired as
\begin{align}\label{equ:ukf92}
&\bar{\theta}_{k}^{s}(\zeta)=\bar{\theta}_{k}(\zeta)+ K_k^s[\bar{\theta}_{k}^{s}(\zeta+1) - {{y}}_k^ {s-} ],\\
& P_{k}^{s}(\zeta)= {{P}}_k(\zeta)+ K_k^{s}[P_{k}^{s}(\zeta+1)-{{P}}_{yy}^{s}]{K_k^{s}}^T.
\end{align}

\begin{remark}
The most computationally expensive operation of UKF and URTSS is to form the sigma points in Eq. \eqref{sigmapoints}, \eqref{gettingsigmapointspredition} and \eqref{sigmapoints2}, which needs the Cholesky decomposition of the state covariance matrices. The complexity can be decreased by intaking the square-root UKF and the square-root URTSS, which propagates the Cholesky factors rather than the state covariance matrices to avoid the decomposition.
\end{remark}
\subsection{ AS $\Delta \theta_k(\zeta)$ Tracking}
 Let us take the first order Taylor series expansion of $\mathbf a[\theta_{kp}(\zeta)]$ around central DOA of $\bar\theta_k(\zeta)$  as
    \begin{align}\label{equ:Taylor}
    \begin{aligned}
         \mathbf a[\theta_{kp}(\zeta)]&=\mathbf a[\bar{\theta}_{k}(\zeta)+\tilde{\theta}_{kp}(\zeta)]\approx \mathbf a[\bar{\theta}_{k}(\zeta)]+\tilde{\theta}_{kp}(\zeta)\frac{\partial\mathbf a[\bar{\theta}_{k}(\zeta)]}{\partial\bar{\theta}_{k}(\zeta)},
    \end{aligned}
    \end{align}
where the high order terms of the series are neglected since the value of $\tilde{\theta}_{kp}(\zeta)$ is small for massive MIMO systems \cite{2D-ESPRIT}.

The uplink received signal can be rewritten as
\begin{align}\label{equ:data received}
\mathbf x(n)=\sum_{k=1}^K\left[\mathbf a[\bar{\theta}_{k}(\zeta)]z_{k1}(n) + \frac{\partial\mathbf a[\bar{\theta}_{k}(\zeta)]}{\partial\bar{\theta}_{k}(\zeta)}z_{k2}(n)\right]+\mathbf w(n),
\end{align}
where $z_{k1}(n)=s_k(n)\sum_{p=1}^Pe^{\vartheta_{kp}}\alpha_{kp}$,
and $z_{k2}(n)=s_k(n)\sum_{p=1}^P e^{\vartheta_{kp}}\alpha_{kp}\tilde {\theta}_{kp}(n)$.
Moreover, it can be computed that $\text{E}\left[z_{k1}(n)z_{k1}^H(n)\right]=S_k\sigma_{k\alpha}^2$,
$\text{E}\left[z_{k2}(n)z_{k2}^H(n)\right]=S_k\sigma_k^2\sigma_{k\alpha}^2$, where $\sigma_{k\alpha}$ is the covariance of $\alpha_{kp}$ and $S_k=\text{E}\left|s_k(n)^2\right|$.

Define\begin{align}
&\mathbf A=\left[ \mathbf a[\bar{\theta}_{1}(\zeta)],\dots, \mathbf a[\bar{\theta}_{K}(\zeta)], \frac{\partial\mathbf a[\bar{\theta}_{1}(\zeta)]}{\partial\bar{\theta}_{1}(\zeta)},\dots,\frac{\partial\mathbf a[\bar{\theta}_{K}(\zeta)]}{\partial\bar{\theta}_{K}(\zeta)}\right],\\
&\mathbf z(n)=\left[z_{11}(n),\dots,z_{K1}(n),z_{12}(n),\dots,z_{K2}(n) \right]^T.
\end{align}
The uplink received signal can be rewritten as
\begin{equation}\label{equ:data received Final}
\mathbf x(n)=\mathbf A\mathbf z(n)+\mathbf w(n).
\end{equation}
The covariance matrix in each block $\zeta$ can be computed as
\begin{equation}\label{equ:covariance matrix}
\mathbf R_{\mathbf x}=\text{E}\left[\mathbf x(n)\mathbf x^H(n)\right]=\mathbf A\mathbf \Sigma \mathbf A^H+\sigma_n^2\mathbf I_M,\ \ n=0,\dots,N-1,
\end{equation}
where $\mathbf \Sigma= \rho \text{diag}( S_1\sigma_{1\alpha}^2,\dots,S_K\sigma_{K\alpha}^2,S_1\sigma_1^2\sigma_{1\alpha}^2,\dots,S_1\sigma_K^2\sigma_{K\alpha}^2)$.

From (\ref{equ:covariance matrix}),  $\mathbf \Sigma$  can be obtained as
\begin{equation}\label{equ:lambda}
\hat{\mathbf \Sigma}=\hat{\mathbf A}^{\dag}(\mathbf R_{\mathbf x}-{\sigma}_n^2\mathbf I_M)(\hat{\mathbf A}^H)^{\dag},
\end{equation}
where $\hat{\mathbf A}$ is the estimation of $\mathbf A$, which can be obtained by replacing the DOA with the tracked~${\bar{\theta}}_k$.

According to the matrix structure of $\hat{\mathbf \Sigma}$, the variance $\sigma_{k}^2$ can be estimated as
\begin{equation}\label{equ:as}
\hat{\sigma}_{k}^2={\frac{[\hat{\mathbf \Sigma}]_{K+k,K+k}}{[\hat{\mathbf \Sigma}]_{k,k}}}.
\end{equation}

Since ${\theta}_{kp}$ is uniformly distributed within $\left[{\bar\theta}_k(\zeta)-\Delta{\theta}_k(\zeta), {\bar\theta}_k(\zeta)+\Delta{\theta}_k(\zeta)\right]$,
we can compute $\Delta{\theta}_k(\zeta)=\sqrt{3\hat{\sigma}_k^2}$ for the current block. Then, according to \eqref{equ:multi-rayrange}, the SSI set of $\mathcal B_k(\zeta)$ for the uplink channel can be derived as
\begin{equation}\label{equ:bk}
\mathcal B_k(\zeta)=\left[\lfloor M\frac{d}{\lambda}\sin\left({\bar\theta}_k(\zeta)-\Delta{\theta}_k(\zeta)\right)\rfloor, \lceil M\frac{d}{\lambda}\sin\left({\bar\theta}_k(\zeta)+\Delta{\theta}_k(\zeta)\right)\rceil\right].
\end{equation}

\begin{remark}
For different deployment circumstances, we can  assume other system equations. For example, we can consider the fixed trajectories for the high-speed railway and unmanned aerial vehicle (UAV) applications. This new information could definitely be utilized to improve the tracking accuracy.
\end{remark}

\section{Tracking the Channel Gains }\label{sec:5}
In this section, we will use a small amount of pilot symbols to track the time-varying channel gains for  both uplink and downlink channels with ST-BEM.

\subsection{ Tracking Uplink Channel Coefficients}
The pilot symbol aided modulation  technique \cite{PSAM} is used to derive the time-varying channels, where the pilots are inserted among the information symbols in each interval of $NT_s$. Define $\mathcal{T}_t=\{n_0,n_1,\ldots,n_{T-1}\}\subset\{0,\ldots,N-1\}$
as the time index set for pilot symbols. Since channels of different users in the same group could be distinguished by their spatial
signatures, we could assign the same pilot sequence to  each user in the same group to save
the orthogonal training resources.
Denote the received training sequences at BS as
$\mathbf Y=[\mathbf y(n_0),\mathbf y(n_1),\ldots, \mathbf y(n_{T-1})]$,
and the common pilot sequences for the $g$-th group
as $\mathbf S_g=\textup{diag}\{s_g(n_0),s_g(n_1),\ldots,s_g(n_{T-1})\}$ with $\sum_{i=0}^{T-1}|s_g(n_i)|^2=1$. Then, it holds that
\begin{align}\label{equ:ULtraining}
\mathbf Y&=\sum_{g=1}^G\sum_{k\in\mathcal{U}_g}
 \left[\mathbf h_k(n_0), {\ldots},\mathbf h_k(n_{T\!-\!1})\right]\sqrt{P_{k}^{\textup{ul}}}\mathbf S_{g}+\mathbf N=\sum_{g=1}^G\sum_{k\in\mathcal{U}_g} \sqrt{P_{k}^{\textup{ul}}}\mathbf F^H\bm \Gamma_k \mathbf C \mathbf S_g +\mathbf N\notag\\
  &=\mathbf F^H\left(\sum_{k\in\mathcal{U}_1}\!\!\sqrt{P_{k}^{\textup{ul}}}\bm \Gamma_k,{\ldots},\sum_{k\in\mathcal{U}_G}\!\!\sqrt{P_{k}^{\textup{ul}}}\bm \Gamma_k\right)  \left[(\mathbf C\mathbf S_1)^H,{\ldots}, (\mathbf C\mathbf S_G)^H\right]^H+\mathbf N\notag\\
  &=\mathbf F^H\bm \Gamma \left[(\mathbf C\mathbf S_1)^H,{\ldots}, (\mathbf C\mathbf S_G)^H\right]^H+\mathbf N,
\end{align}
where $P_{k}^{\textup{ul}}$ is the uplink power constraint of user $k$, $\bm \Gamma_k=[\bm\gamma_{k,0},\bm \gamma_{k,1},\ldots,\bm \gamma_{k,M-1}]^T$ denotes the CE-BEM coefficients for user-$k$, $\mathbf C=[\mathbf c_{n_0},\mathbf c_{n_1},\ldots,\mathbf c_{n_{T-1}}]$,
 and $\mathbf N$ is the noise matrix whose elements are  i.i.d. $\mathcal{CN}(0,\sigma_n^2)$.

When $T\geq G(\mu+1)$, there will be adequate observations to estimate all unknowns parameters in $\bm \Gamma$, and the standard least square (LS) estimator yields \cite{gao_relay}
\begin{align}
  \hat{\bm \Gamma} = \mathbf F\mathbf Y\left\{\left[(\mathbf C\mathbf S_1)^H,{\ldots}, (\mathbf C\mathbf S_G)^H\right]^{H}\right\}^{\dag}.
\end{align}
The mean square error (MSE) of $\hat{\bm \Gamma}$ can be computed as
\begin{align}\label{equ:ULMSE}
  \mathbb E\{\|\hat{\bm \Gamma}-\bm \Gamma\|_F^2\}=M\sigma_n^2\ \textup{tr}\left\{\left(\left[(\mathbf C\mathbf S_1)^H,{\ldots}, (\mathbf C\mathbf S_K)^H\right]^H\left[\mathbf C\mathbf S_1,{\ldots},\mathbf C\mathbf S_K\right]\right)^{-1}\right\}.
\end{align}

According to \cite{gao_relay}, we know that
the optimal pilot symbols to minimize the MSE in \eqref{equ:ULMSE} should satisfy the following constraints.
\begin{align}\label{equ:ULoptimaldesign}
  \mathbf C\mathbf S_g\mathbf S_g^H\mathbf C^H=\mathbf I_{\mu+1},\
  \mathbf C\mathbf S_g\mathbf S_{g'}^H\mathbf C^H=\mathbf 0_{\mu+1}, \ \forall\ g\neq g'.
\end{align}
Then, the optimal pilot symbols for different groups are proved to be equi-powered, equi-spaced over $\{0,\ldots,N-1\}$,
and should be phase shift orthogonal  \cite{gao_relay}. One typical example of this kind of
pilot sequence is
\begin{align}\label{equ:optimalsequence}
  s_g(n_i)=\sqrt{1/T}e^{j2\pi i (g-1)(\mu+1)/T},\notag g=1,\ldots,G,\  i=0,\ldots,T-1.
\end{align}

Next, we focus on users in group $g$ and derive that
\begin{align}
  \hat{\bm \Gamma}_k = \bm \Gamma_k+\sum_{l\in\{\mathcal{U}_g\backslash k\}}\sqrt{P_{l}^{\textup{ul}}/P_{k}^{\textup{ul}}}\bm \Gamma_l+
  \frac{1}{\sqrt{P_{k}^{\textup{ul}}/\sigma_n^2}}\mathbf N_k,
\end{align}
where $\mathbf N_k{\in}\mathbb{C}^{M\times \mu+1}$ has the i.i.d. $\mathcal{CN}(0,1)$ elements.

Considering the distinct spatial signatures of users in the same group, we can  extract
\begin{align}\label{equ:ULlambda_hat}
   \widehat{\left[\bm \Gamma_k\right]}_{\mathcal{B}_k(\zeta),:}=[\hat{\bm \Gamma}_k ]_{\mathcal{B}_k(\zeta),:} =\left[\bm \Gamma_k\right]_{\mathcal{B}_k(\zeta),:}
+\sum_{l\in\{\mathcal{U}_g\backslash k\}}\sqrt{P_{l}^{\textup{ul}}/P_{k}^{\textup{ul}}}\left[\bm \Gamma_l\right]_{\mathcal{B}_k(\zeta),:}
+\frac{\left[\mathbf N_k\right]_{\mathcal{B}_k(\zeta),:}}{\sqrt{P_{k}^{\textup{ul}}/\sigma_n^2}},
\end{align}
where the second term $\sum_{l\in\{\mathcal{U}_g\backslash k\}}\sqrt{P_{l}^{\textup{ul}}/P_{k}^{\textup{ul}}}\left[\bm \Gamma_l\right]_{\mathcal{B}_k(\zeta),:}$ is the pilot contamination term caused by reusing the same pilot in one group. Since $\mathcal{B}_l(\zeta)$ and $\mathcal{B}_k(\zeta)$ are kept away from each other, the entries of $\left[\bm \Gamma_l\right]_{\mathcal{B}_k(\zeta),:}$ in \eqref{equ:ULlambda_hat} are very small and negligible. Therefore, $\bm \Gamma_k$ can be approximated as
\begin{align}\label{equ:coefficient}
  \bm \Gamma_k=\left[\mathbf 0^T\ \widehat{\left[\bm \Gamma_k\right]}_{\mathcal{B}_k(\zeta),:}^H\ \mathbf 0^T\right]^H.
\end{align}

\begin{remark}
While $T=K(\mu+1)$ pilots are needed for traditional channel estimation scheme \cite{xie}, the total pilot overheads of uplink channel tracking is significantly reduced to $T=G(\mu+1)$ ($G\ll K$) through the proposed method, which greatly decreases the uplink training overhead.
\end{remark}

\subsection{ Tracking Downlink Channel Coefficients} \label{subsec:1}
Denote the downlink channel from BS to user-$k$ as $\mathbf g_k(n)\in\mathbb{C}^{M\times 1}$. Similar to (\ref{equ:equal}), $\mathbf g_k(n)\in\mathbb{C}^{M\times 1}$ can be modeled as
 \begin{align}\label{equ:DLSTBEM}
   \mathbf g_k(n)\approx \sum_{q\in \overline{\mathcal{B}_k}} \tilde{g}_{k,q}(n)\mathbf f_q= \sum_{q\in\overline{\mathcal{B}}_k(\zeta)} \bm{\gamma}_{k,q}^{d^T} \mathbf c_n \mathbf f_q,n=0,\dots,N-1,
 \end{align}
where $\bm \gamma_{k,q}^{d}$ denotes the downlink channel CE-BEM coefficients and
$\overline{\mathcal{B}}_k(\zeta)$ is the downlink spatial signatures. Consistent with the uplink channel tracking, the downlink channel tracking can also be simplified to tracking the downlink SSI set $\overline{\mathcal{B}}_k(\zeta)$ and estimating those remaining unknown coefficients $\bm \gamma_{k,q}^{d}$.

Traditionally, in the mode of TDD, the downlink channel state information can be directly derived from channel reciprocity. However, in the
mode of FDD where the channel reciprocity would not hold, the training overhead for the high
dimension channel would be prohibitive, which is the main challenge of massive MIMO
system. Interestingly, electromagnetic characteristics do not change in several dozens of GHz
\cite{DOAreciprocity1 ,DOAreciprocity2,DOAreciprocity3}, and the propagation path of electromagnetic wave is reciprocal, i.e., uplink DOA and downlink DOD are reciprocal. Therefore, we could utilize the angle reciprocity
to simplify the complexity of downlink channel tracking in FDD mode.

With angle reciprocity, the downlink spatial signature, denoted as  $\overline{\mathcal{B}_k} (\zeta)$, can be directly obtained from the tracked uplink SSI set $\mathcal{B}_k(\zeta)$. With the angle reciprocity, there is
\begin{align}
\sin\theta_{kp}(\zeta) = \frac{q\lambda_1}{Md}=\frac{q'\lambda_2}{Md},\ \textup{with}\ q\in\mathcal{B}_k(\zeta),\
q'\in\overline{\mathcal{B}}_k(\zeta),
\end{align}
where $\lambda_1$ and $\lambda_2$ denote the uplink and downlink carrier wavelength respectively.
Then the downlink SSI set $\overline{\mathcal{B}_k}(\zeta)$ can
be expressed as
\begin{align}\label{equ:DLBk}
\overline{\mathcal{B}}_k(\zeta)=\left[\left\lfloor\frac{\lambda_1}{\lambda_2}q_{\min}\right\rfloor,\left\lceil\frac{\lambda_1}{\lambda_2}q_{\max}\right\rceil\right],
\end{align}
where $q_{\min}\leq q\leq q_{\max}, \forall q\in\mathcal{B}_k(\zeta)$.

Denote the size of $\overline{\mathcal{B}}_k(\zeta)$ as $\tau$, i.e., the effective dimension of downlink channel.  According to \eqref{equ:DLSTBEM}, $T=\tau(\mu+1)$ pilot symbols are required for downlink channel parameter estimation. Adopting the uplink user grouping strategy directly, the received $\mathbf y^d_k=[y_k(n_0),\ldots,y_k(n_{T-1})]^T$ at user-$k$ in $\mathcal{U}_g$ can be expressed as
\begin{align}\label{equ:DLtraining}
  \mathbf y^d_k=&\sum_{l\in\mathcal{U}_g}\sum_{1\leq i\leq \tau, q_i\in\overline{\mathcal{B}}_l(\zeta)}\sqrt{P_l^{\textup{dl}}/\tau}\mathbf S^d_i\mathbf C^T\bm\gamma_{k,q_i}^d\mathbf f_{q_i}+\mathbf n_k\notag\\
  =&\sum_{l\in\mathcal{U}_g}\left[\sqrt{P_1^{\textup{dl}}/\tau}\mathbf S^d_1\mathbf C^T,\ldots,\sqrt{P_\tau^{\textup{dl}}/\tau}\mathbf S^d_{\tau}\mathbf C^T\right]
  \textup{vec}\left(\left[\bm\Gamma_k^{d}\right]_{\overline{\mathcal{B}}_l(\zeta)}^T\right)
  \mathbf F_{\overline{\mathcal{B}_l(\zeta)},:}+\mathbf n_k,
\end{align}
where $\bm\Gamma_k^{d}$ is the downlink channel parameters, $\mathbf n_k\sim\mathcal{CN}(\mathbf 0,\sigma_n^2\mathbf I_T)$ is the noise vector, $P_l^{\textup{dl}}$ is the total downlink power constraint for user $l$ and diagonal matrices $\mathbf S^d_i\in\mathbb{C}^{T\times T},i=1,\ldots,\tau$ denote the transmitted pilot sequences for users in $\mathcal{U}_g$.

\begin{figure}[t]
\centering
\includegraphics[width=90mm]{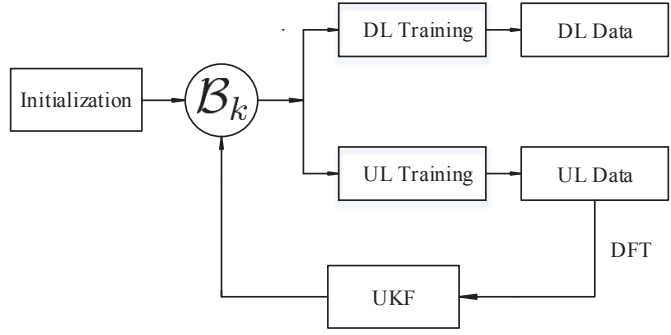}
\caption{The overall channel tracking strategy, where `DL' means the downlink channel and  `UL' means the uplink channel. The initial SSI sets $\mathcal B_k(1)$ can be obtained from the method in \cite{xie}.
\label{fig:channeltracking}}
\end{figure}

Similar to \eqref{equ:optimalsequence}, the optimal pilot sequences $\mathbf S^d_i\in\mathbb{C}^{T\times T},i=1,\ldots,\tau$
for downlink channel tracking are also equi-powered, equi-spaced and
phase shift orthogonal. Then, the downlink channels can be recovered by
\begin{equation}\label{equ:DLtraining1}
\widehat{\left[\bm \Gamma_k^d\right]}_{\overline{\mathcal{B}}_k(\zeta)}=\left[\bm \Gamma_k^d\right]_{\overline{\mathcal{B}}_k(\zeta)}
+\!\!\sum_{l\in\{\mathcal{U}_g\backslash k\}}\sqrt{P_{l}^{\textup{dl}}/P_{k}^{\textup{dl}}}\left[\bm \Gamma_l^d\right]_{\overline{\mathcal{B}}_k(\zeta)}
+\frac{1}{\sqrt{P_{k}^{\textup{dl}}/\sigma_n^2}}\mathbf n_k^{\prime},%
\end{equation}
where $\mathbf n_k^{\prime}{\in}\mathbb{C}^{\tau\times 1}$ has the i.i.d. $\mathcal{CN}(0,1)$ elements.

It can be found that user $k$ does not need the knowledge of spatial SSI set ${\overline{\mathcal{B}}_k(\zeta)}$ to perform the estimation of $\widehat{[\bm \Gamma_k^d]}_{\overline{\mathcal{B}}_k(\zeta)}$. Each user only needs to feed back $\tau$ components $\widehat{[\bm \Gamma_k^d]}_{\overline{\mathcal{B}}_l(\zeta)}$
to BS to fulfill the downlink channel tracking, which removes the necessity of feedback from BS to the user. This is a key advantage that makes the proposed downlink channel tracking strategy suitable for fast-fading environments. The overall channel tracking mechanism is illustrated  in Fig.~\ref{fig:channeltracking}.

\section{Simulations }\label{sec:6}
\begin{figure}[t]
\centering
\includegraphics[width=120mm]{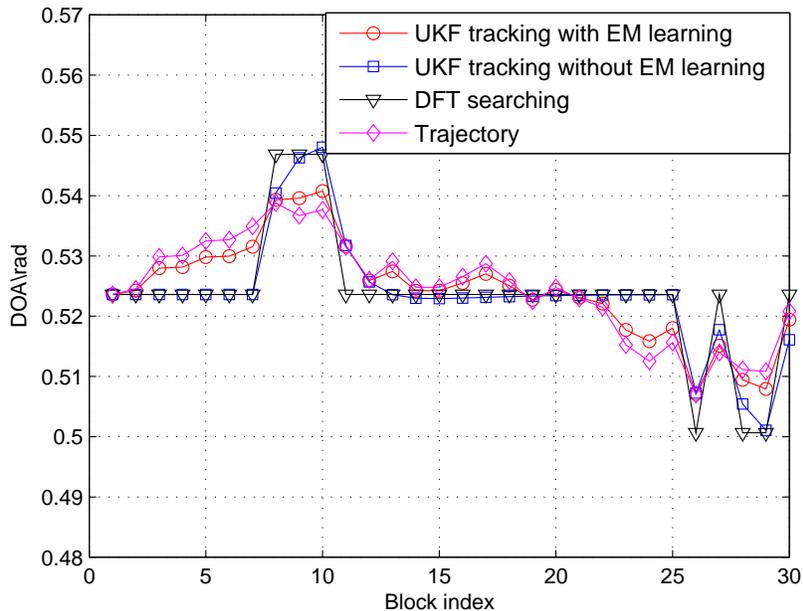}
\caption{Performance of central DOA tracking with SNR $\rho=10$ dB.}
\label{fig:track}
\end{figure}

\begin{figure}[t]
\centering
\includegraphics[width=110mm]{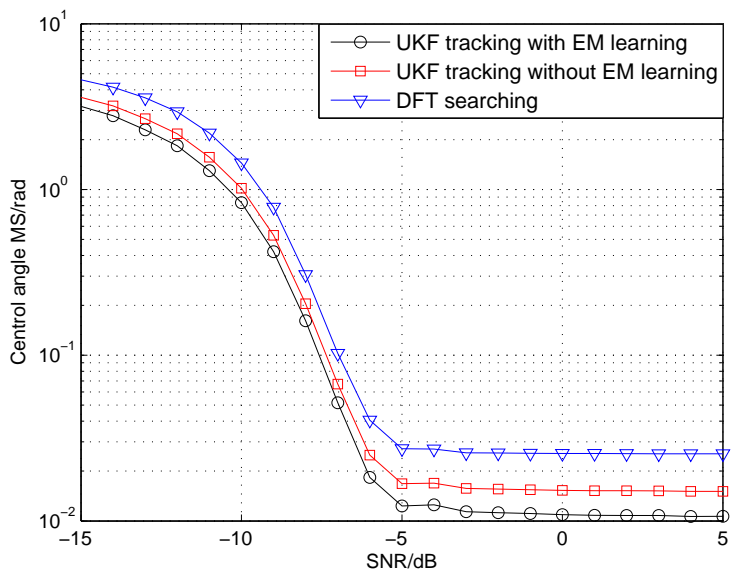}
\caption{Performance of central DOA tracking with different SNR.}
\label{fig:msedoa}
\end{figure}

 In this section, numerical results are presented to demonstrate the effectiveness of the proposed methods. We set $M=128$,  $d=\frac{\lambda}{2}$ and  $K=12$. Besides, all $K$ users are gathered into 4 spatially distributed clusters, and users can be divided into $G=3$ groups according to the DOA distribution. The channel vector between the users and the BS is generated according to (\ref{equ:channelmodel}) with AS $\Delta \theta=2^\circ$.
 The performance metric of the channel estimation is taken as the normalized MSE,~i.e.
\begin{align}
  \tu{MSE} \triangleq  \frac{1}{N}\sum_{n=1}^N\sum_{k=1}^K\frac{\left\|\mathbf h_k(n)-\hat{\mathbf h}_k(n)\right\|^2}{\left\|\mathbf h_k(n)\right\|^2}.
\end{align}

In the first example, the performance of the central DOA tracking is shown in Fig. \ref{fig:track}, where the real trajectory, central DOA from DFT searching, and central DOA tracking without EM leaning are also displayed for comparison. Meanwhile, the DOA tracking performance as a function of SNR is also displayed in Fig. \ref{fig:msedoa}. The users moves with maximum speed 80 Km/h, and the variation of DOA is generated according to the system equation \eqref{equ:kinematic}. It can be found from Fig. \ref{fig:track} that the angle variation is small for massive MIMO systems, and both the DFT searching method and the central DOA tracking without EM learning are approximately consistent with the trajectory but have larger error than the DOA tracking with EM learning. The reason is that the one-order Markov process can well describe the movements of the users, and EM learning can precisely derive the unknown parameters from the system equation and measurement equation. The MSE performance of central DOA tracking with different SNR in Fig. \ref{fig:msedoa} further verifies the aforementioned analysis. Besides, we can also find that there are error floors for all the displayed method, which arises from the observation error.

\begin{figure}[t]
\centering
\includegraphics[width=110mm]{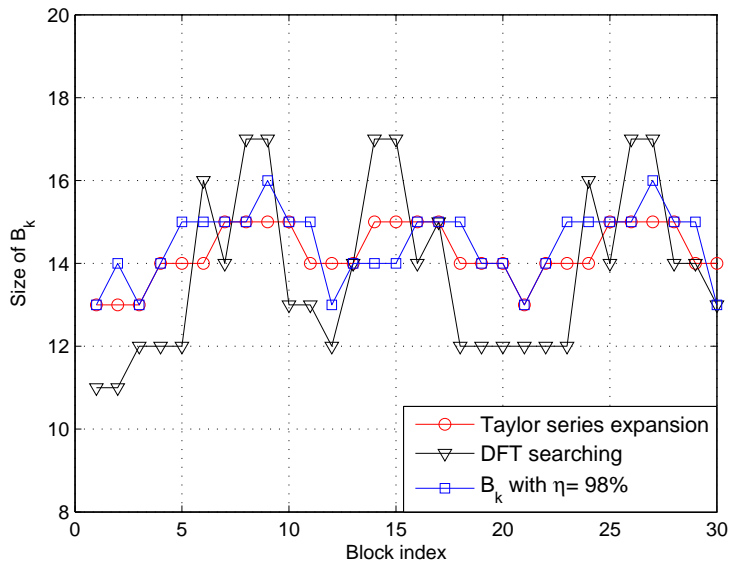}
\caption{Performance comparison of AS tracking with SNR $\rho=10$ dB.
\label{fig:B}}
\end{figure}

Fig. \ref{fig:B} displays the performances of AS tracking from DFT searching and Taylor series expansion, respectively. Meanwhile, the reference curve, which represents $\mathcal B_k(\zeta)$ that contains $98\%$ power of the channel, is also displayed as comparison. It is seen that the result from DFT searching is not desirable and yields large fluctuation errors, while the one from Taylor series expansion method is much better and closer to the reference curve. The reason is that when the number of the antennas $M$ is not infinite, the power leakage effect and the measurement noise will affect the precision of SSI set tracking. 

\begin{figure}[t]
\centering
\includegraphics[width=110mm]{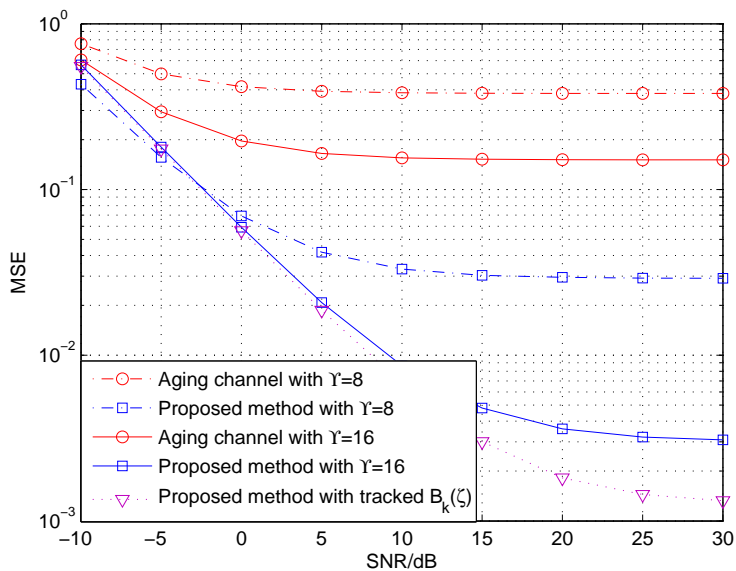}
\caption{Uplink MSE performance of the proposed method.
\label{fig:mse}}
\end{figure}

Fig. \ref{fig:mse} displays the uplink channel tracking performances as a function of SNR and MSE, where the aging channel without spatial information tracking and the channel with tracked $\mathcal B_k(\zeta)$ are also displayed as comparison. Besides, since the communications standards normally regulate a constant number of system parameters rather than a dynamic quantity for practical deployments, the performances of the proposed method with a predefined size of SSI set, denoted as $\Upsilon$, are also displayed as comparison. It can be seen that as the SNR increases, there are error floors for the ST-BEM based methods. This phenomenon arises from the truncation error of BEM from the real channel and can also be observed in \cite{GiannakisBEM}. Besides, we can also see that the performance of channel tracking behaves better with the increase of $\Upsilon$, and the MSE of the proposed method with the tracked $\mathcal B_k(\zeta)$ is the best in the displayed curves.

\begin{figure}[t]
\centering
\includegraphics[width=110mm]{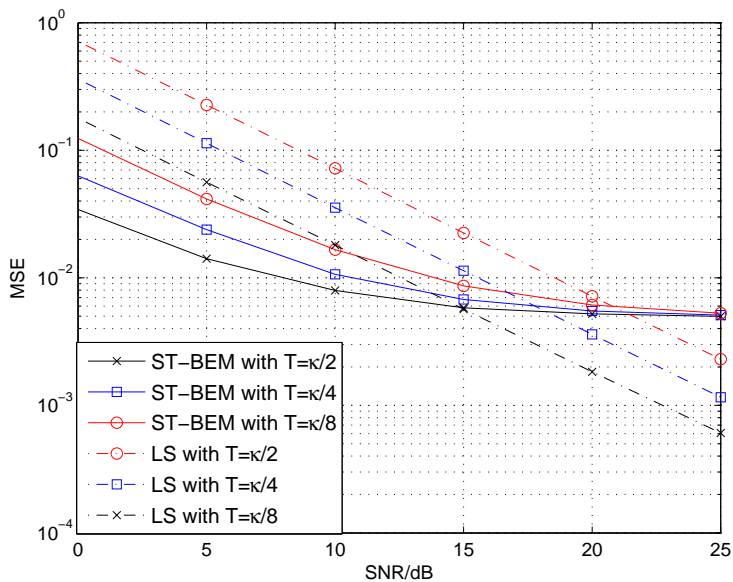}
\caption{Downlink MSE performance comparison of ST-BEM with $T=\kappa/8, \kappa/4, \kappa/2$, respectively, and conventional LS with $\kappa =M(\mu+1)$.
\label{fig:DLMSE_T}}
\end{figure}

Fig. \ref{fig:DLMSE_T} compares the downlink MSE performance of the proposed method, where $\kappa =M(\mu+1)$ pilot symbols are utilized for the conventional LS method, while  $T{=}\kappa/8, \kappa/4, \kappa/2$ are allocated for the ST-BEM method separately. The power constraints are the same $\sum_{k=1}^KP^{\textup{dl}}_k=KT\rho$ for both methods for fairness. It can be found that the proposed method is much superior to the conventional LS method, not only for the higher estimation accuracy in low SNR regions, but also due to the less training overheads, which is consistent with the analysis in the Section \ref{subsec:1}.

\begin{figure}[t]
\centering
\includegraphics[width=110mm]{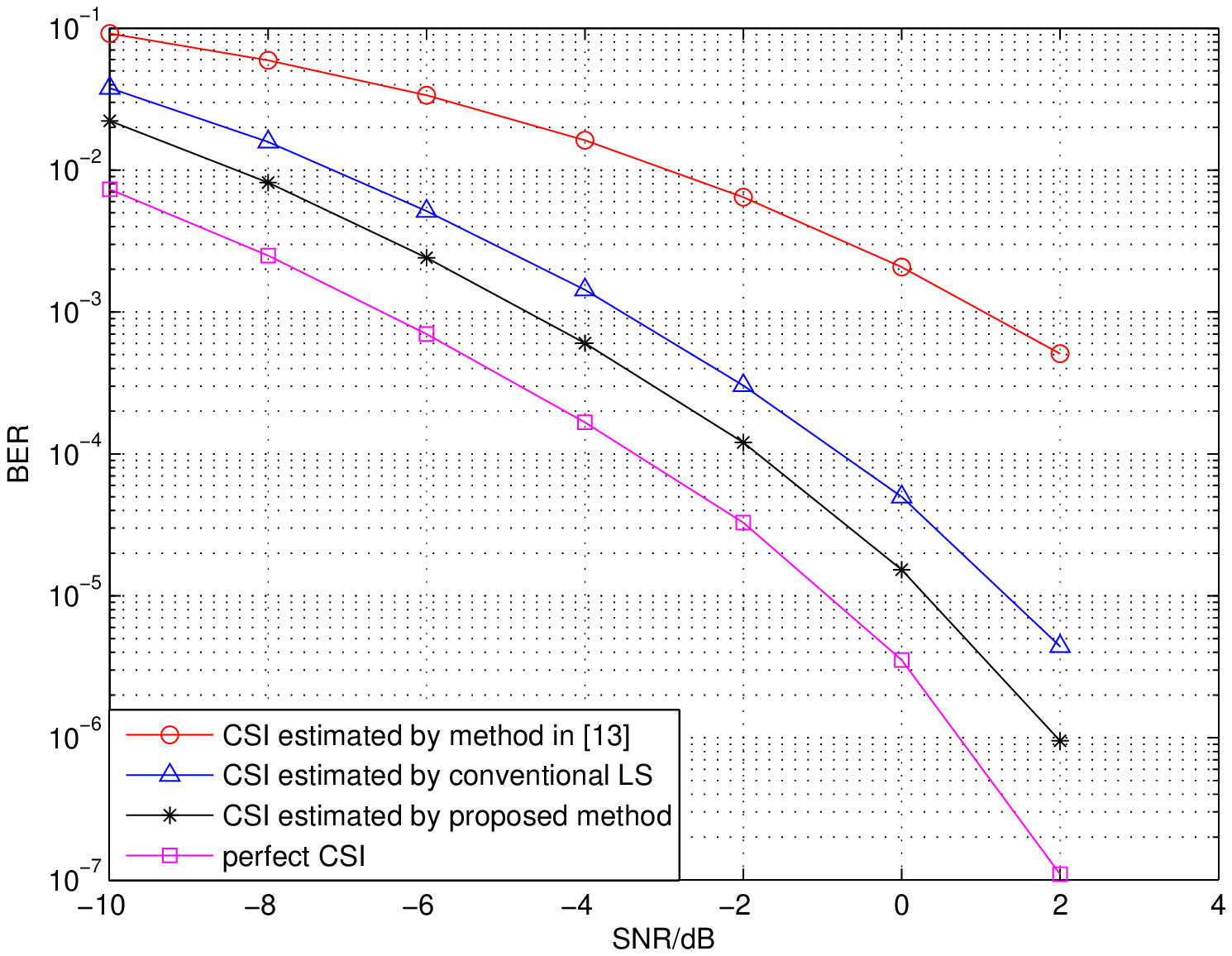}
\caption{BER comparison of perfect CSI, CSI from the proposed method, CSI from the method in \cite{xie}, and CSI from the conventional LS method with the same training power.
\label{fig:BER}}
\end{figure}

Lastly, the bit error rate (BER) performance under QPSK modulation is shown for the downlink data transmission in Fig. \ref{fig:BER}. The results  of perfect CSI, CSI from the proposed method, CSI from the SBEM method \cite{xie}, and CSI from the conventional LS method are displayed under the matched filter precoding scheme. The overall training power is set as the same for each method to keep the comparison fairness. It is easy to find that the method in \cite{xie} is not efficient for highly time-varying channel. The proposed channel tracking method performs better than the LS method at the low SNR region, and the reasons can be found from \eqref{equ:ULlambda_hat} where the proposed method only involves $\tau$ components of the noise vector, while the conventional LS method includes the whole noise power.
Moreover, it can be seen that the BER achieved by ST-BEM has about 0.5 dB gap from that of perfect CSI, which corroborates the effectiveness of the proposed ST-BEM channel tracking method.

\section{Conclusion}\label{sec:7}
In this paper, the problem of time-varying channel tracking was considered for massive MIMO systems. We proposed an ST-BEM to reduce the effective dimensions of the channels, where the spatial channel is decomposed into the time-varying spatial information and the time-varying gain information. Since the spatial information is characterized by the central DOA and AS of the incident signal, we use the one-order Markov process to describe the user's movements and UKF to blindly track the central DOAs, where the unknown system parameters in the system equation and measurement equation are estimated by EM learning. In addition, AS is also blindly tracked through Taylor series expansion of the steering vector. With the derived SSI set, the time-varying gain information of ST-BEM can be derived through a few pilot symbols. Simulation results show that the proposed method offers a practical and effective way for massive MIMO channel tracking.


\ifCLASSOPTIONcaptionsoff
\fi
\linespread{1.5}

\end{document}